# Wavelet Based Volatility Clustering Estimation of Foreign Exchange Rates

A.N.Sekar Iyengar,

Saha Institute of Nuclear Physics,

1/AF Bidhan Nagar, Kolkata-700 064

ansekar.iyengar@saha.ac.in

**Wavelet Based Volatility Clustering Estimation of Foreign Exchange Rates**

## Abstract

We have presented a novel technique of detecting intermittencies in a financial time series of the foreign exchange rate data of U.S.- Euro dollar( US/EUR) using a combination of both statistical and spectral techniques. This has been possible due to Continuous Wavelet Transform (CWT) analysis which has been popularly applied to fluctuating data in various fields science and engineering and is also being tried out in finance and economics. We have been able to qualitatively identify the presence of nonlinearity and chaos in the time series of the foreign exchange rates for US/EURO (United States dollar to Euro Dollar) and US/UK ( United States dolar to United Kingdom Pound) currencies. Interestingly we find that for the US-INDIA( United States dollar to Indian Rupee) foreign exchange rates, no such chaotic dynamics is observed. This could be a result of the government control over the foreign exchange rates, instead of the market controlling them.

(**key words**: Time-Scale analysis, Intermittency, Nonlinearity and Chaos

## Extended Summary

A significant number of physicists are engaged in the investigation of the behaviour of the financial markets with a view to understanding its behaviour and to model and forecast its future behaviour. Amongst this nonlinear dynamics is seen to play a significant role in the evolution of the financial markets and it is certain that it will play a bigger role in modeling the financial data in the future. Prior to even applying the nonlinear dynamical techniques it is essential to know whether there is a nonlinearity involved in these data sets. Most of the time this is done by analyzing the financial time series and estimating some exponents like correlation dimension and the lyapunov exponents. Many at times it may be misleading only to estimate the exponents since two time series may look the same. So one has to rely on other methods as well. In this paper we have analysed three foreign exchange rates using wavelet analysis and have been able to identify the presence of nonlinearity in them. In cases where the financial time series is controlled by the market dynamics it is clearly seen that nonlinearity is very clearly identified and it turns out to be strongly chaotic whereas in some cases where the market forces do not control, but the government controls the evolution of the financial markets, the time series does not exhibit any nonlinearity at all.

**Biographical sketch**

The author is a senior professor of physics at the Saha Institute of Nuclear Physics. His main research interests is in nonlinear aspects of plasma physics and presently is involved in applying the techniques of nonlinear dynamics to understand the instabilities and turbulence in physical, biological and environmental systems. He has published more than 30 papers in refereed journals and guided four Ph.D scholars.

## 1. INTRODUCTION

Financial markets are perhaps the most complex of all systems involving a large number of human beings, with psychological, cultural, social factors also playing a significant role. Cross disciplinary studies on financial systems has attracted much attention recently and this has led to the birth of a new branch of physics called "Econophysics". (Mantegna and Stanley 1995, 1996, Ghasgaie et al 1996, Sitabhra Sinha and Bikas K.Chakrabarti 2009).

Intermittency is a fundamental dynamical feature of complex economic systems. Financial returns are known to be non-Gaussian and exhibit fat-tailed distribution, which relate to intermittent occurrences of large bursts (volatility clustering) -an unexpected high probability of price changes- important for risk analysis. The recent development of high-frequency data bases makes it possible to study the intermittent market dynamics of time scales of less than a day. The intermittent behaviour has been attributed to nonlinear affects. Vassilicos et al ( Vassilicos et al 1993) have shown that the financial time series of foreign exchange rates exhibit multifractality but not chaos while some believe that it could be chaotic (Krawieckij et al 2002).

In order to carry out an efficient modeling and prediction one should have a reliable technique and no single technique can elucidate the entire dynamics that leads to the generation of the financial time series. In this paper we have extensively analyzed the foreign exchange rates of US/EURO, US/UK, and US/INDIA currencies obtained from http://www.federalreserve.gov. Using continuous wavelet transforms we have been able to apply both statistical and spectral analysis to the above data simultaneously. We have also qualitatively been able to locate presence of nonlinearity and chaos in the system.

## 2. METHODOLOGY

### 2.1. Wavelet Analysis

Wavelet analysis though a very old mathematical topic (Weierstrass(1895), Haar(1910) has become a popular tool to analyze complex signals only in the last two or three decades. Its application was first attempted by Morlet and Grossman to analyze Seismic data ( Morlet et al 1982), Gopuillaud et all 1984)) who laid the mathematical foundation for wavelets in 1984 ( Grossman and Morlet 1984). Since then the subject has developed in leaps and bounds with its application encompassing numerous fields like physics, chemistry, medicine, biology, finance etc. A significant advantage is that it can also be used for statistical data analysis in addition to its competence in spectral analysis.

Wavelets were used in fluid mechanics to (Farge 1992, 1996) to extract information about the turbulent or eddy structures. A significant advantage is that they have the potential of extracting the features of structures in flow fields which one can miss out by traditional statistical methods. In this paper we will apply wavelet transform to analyze presence of coherent structures (which are supposed to be responsible for intermittency (volatility clustering)). Though it has been recognized that most financial markets are nonlinear complex systems there is still an unresolved controversy regarding the chaoticity of these markets. It was shown by Vassilicos et al ( Vassilicos et al 93) that there is no chaos in the foreign exchange market. Their work was done during a period when wavelets were not yet exploited to its full capacity. It was shown by Rick Lind et al (2001) that continuous wavelet transform can be used to qualitatively exhibit the presence of nonlinearity. Later Chandre et al ( Chandre et al 2003 ) proposed an interesting method to distinguish qualitatively the presence of chaos ( strong, moderate and weak) using wavelets.

Wavelet analysis basically incorporates time resolution in a more fundamental way than is permitted by the Fourier type methods. It has been applied quite successfully to turbulent data. New methods have been suggested to summarize and visualize the large amount of data generated by the wavelet transform and extract the information relevant to nonlinear interaction. The analyzing basis functions are oscillating functions that decay rapidly with time and are termed as wavelets. Thus a wavelet transform at a given time is similar to a Fourier transform in the sense that it exhibits the time contribution of the different frequencies to the signal, but due to the decay of the wavelet this information only pertains to a certain short interval of the signal. As the signal is scanned by the wavelet on successive time intervals, one obtains a temporal frequency dependence. Rather than referring to frequency the wavelet transform is usually expressed in terms of the scale of the analyzing wavelet which can be understood to be proportional to the inverse frequency. A mathematical description of the above can be expressed as follows. The Fourier transform of a function is given as :

$$\Phi(\omega) = \int \varphi(t)\, e^{-i\omega t} \qquad \ldots\ldots\ldots\ldots\ldots..(1)$$

$$P_{\varphi}(\omega) = |\varphi(\omega)|^{\wedge 2} \qquad \ldots\ldots\ldots\ldots\ldots..(2)$$

A wavelet can be any function $\psi(t)$ that satisfies the following admissibility conditions

$$\int \psi(t)\, dt = 0 \qquad \ldots\ldots..........(3)$$

$$\int |(\psi(\omega)|^{\wedge 2} * |(\omega)|^{-1}\, d\omega < \infty \qquad \ldots\ldots\ldots\ldots...(4)$$

The corresponding wavelet family is obtained by means of the scale length parameter a:

$$\psi_a(t) = 1/a^p * \psi(t/a) \qquad \ldots\ldots\ldots\ldots...(5)$$

The factor p is a normalization choice. The wavelet transform of a function $\varphi(t)$ is then given by

$$W_{\varphi}(a,\tau) = \int \varphi(t)\, \psi_a(t-\tau)\, dt. \qquad \ldots\ldots\ldots\ldots.(6)$$

$W_{\varphi}(a,\tau)$, at any given a , can be interpreted as a filtered version of $\varphi(t)$, band-passed by the filter $\psi_a$. By convention for visualization purposes $|W_f(a,\tau)|^{\wedge 2}$ is plotted in the $(a,\tau)$ plane.

In this paper we mainly used the Daubechies (Db) family of wavelets. These wavelets satisfy the vanishing moment conditions and hence the low-pass coefficients keep track of the polynomial trends in the data. For eg. The Db-4, Db-6 and Db-8 retain polynomial trend which are linear, quadratic and cubic respectively. We have used the Db-7 wavelet for our analysis in this paper.

## 3. RESULTS and DISCUSSION

Fig 1a. shows the raw data of the US/EURO dollar exchange rates for about 9 years from 1 Jan 2000 to 3 March 2009 ( top panel), log - returns ( middle panel) obtained after de-trending the raw data and the power spectrum ( lower panel) of the log - returns. It is clearly seen that it is a broad band spectrum indicating stochastic nature. Fig 1b shows the continuous wavelet spectrum of the log-returns ( top most panel), the raw data corresponding, the probability distribution function of the time series and the lowest panel is the kurtosis of the scale 10 time series respectively.

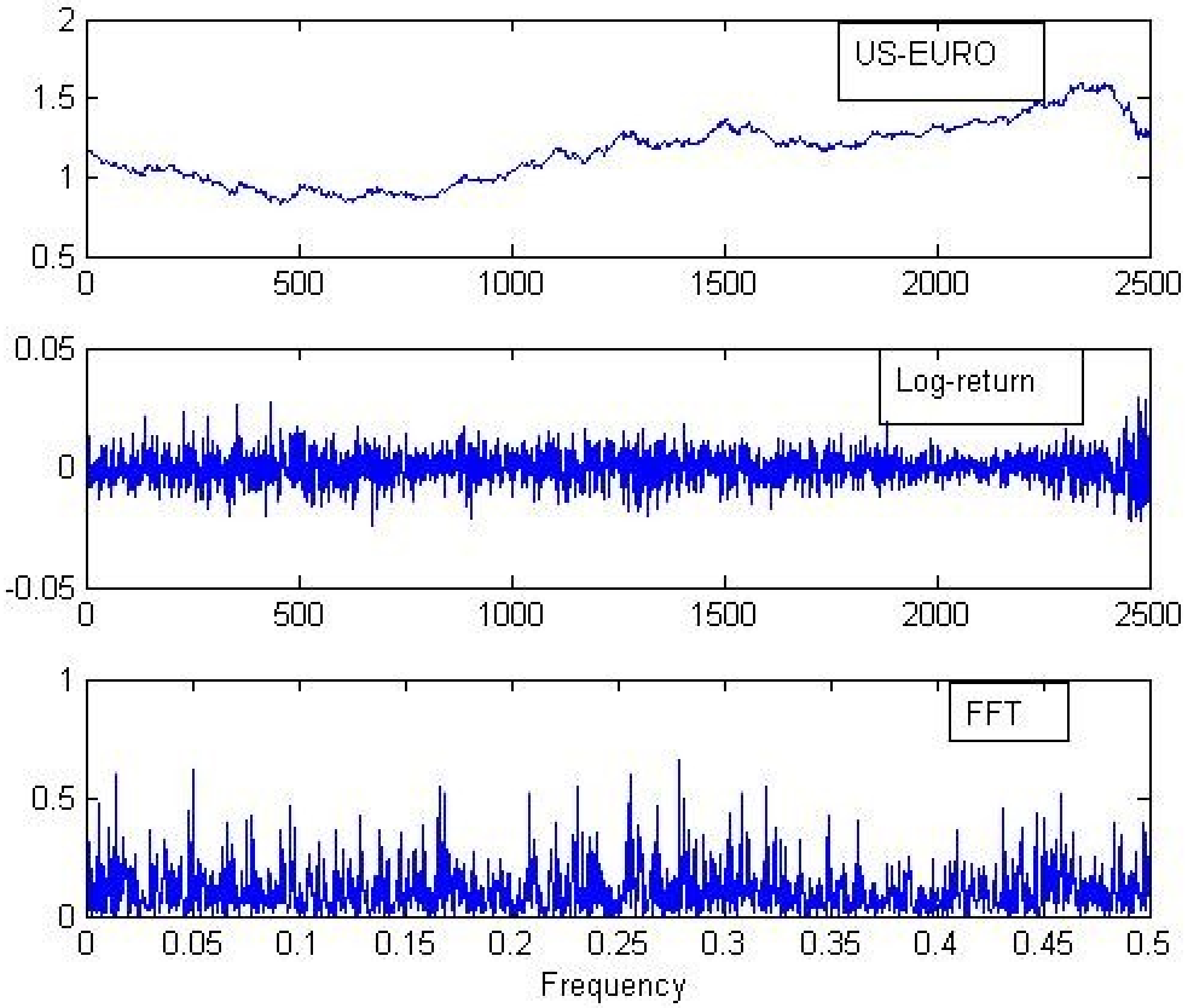

Fig1. a) Raw data of the foreign exchange rate of U.S-Euro dollar ( US-EURO) from 03 March 2000 to 09 March 2009 ( top most panel), log-returns of the raw data ( middle panel) and the Power Spectrum ( lowest panel) of the log-returns which is seen to be of broad band in nature. Top and middle panels ( x-axis is time and y-axis is amplitude in arbitrary units). Lowest panel ( x axis-frequency and y-axis is the power spectral amplitude).

Fig 1b is a two dimensional plot of the time-scale diagram obtained from the continuous wavelet transform. We have used the db-7 (daubechies-7) as the mother wavelet and is clearly seen that there is a significant nonlinear process involved in the generation of the time series and also it is chaotic. Nonlinearity is identified by the presence of the scale shift over several temporal ranges for e.g between 300:400, 800:1100,1200:1300 and so on. The bright coloured regions at the ends may be artifacts. According to Chandre et al. if one were to join the maximum of the contours along the scales we would clearly observe discrete structures and not continous lines indicating the presence of at least moderate chaos. Between the scales of 1 to 30 it is clearly seen that is very strongly chaotic. So one can conclude that nonlinearity and chaos are playing a significant role in the foreign exchange rate dynamics.

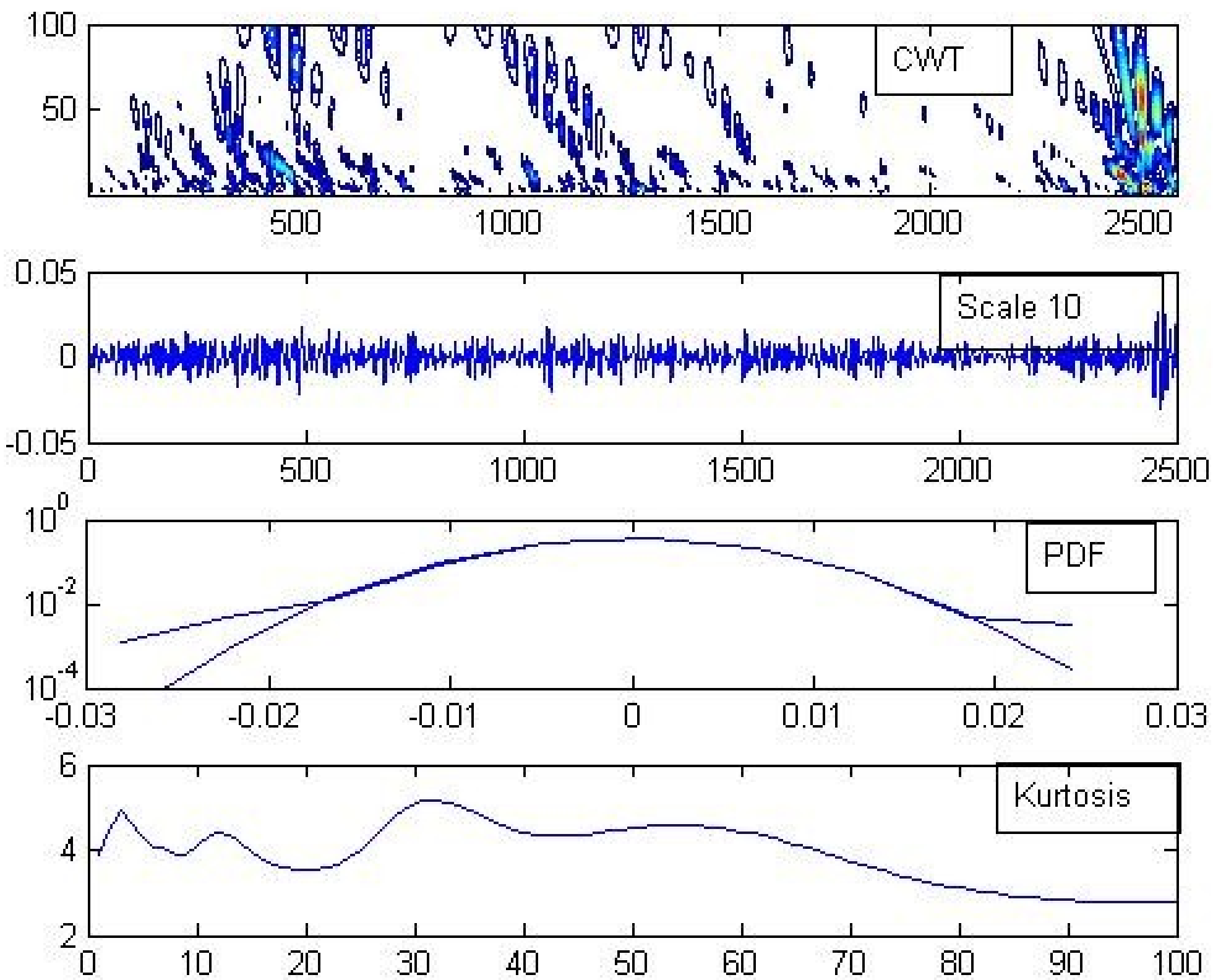

Fig1b. Time-Scale (Time along the x-axis and Scale along the y-axis) diagram obtained from the continuous wavelet transform (Top most panel), time series corresponding to scale 10 ( second from top panel), probability distribution function of the scale 10 time series data ( third panel from top) and kurtosis (lower most panel) of the scale 10 time series. Top and the second from top panels ( x-axis is time and the y-axis is the amplitude in arbitrary units), Third from top panel ( x-axis is the amplitude of the fluctuations and y-axis is the probability amplitude) and the Lowest panel ( x-axis is the scale and the y-axis is the kurtosis value)

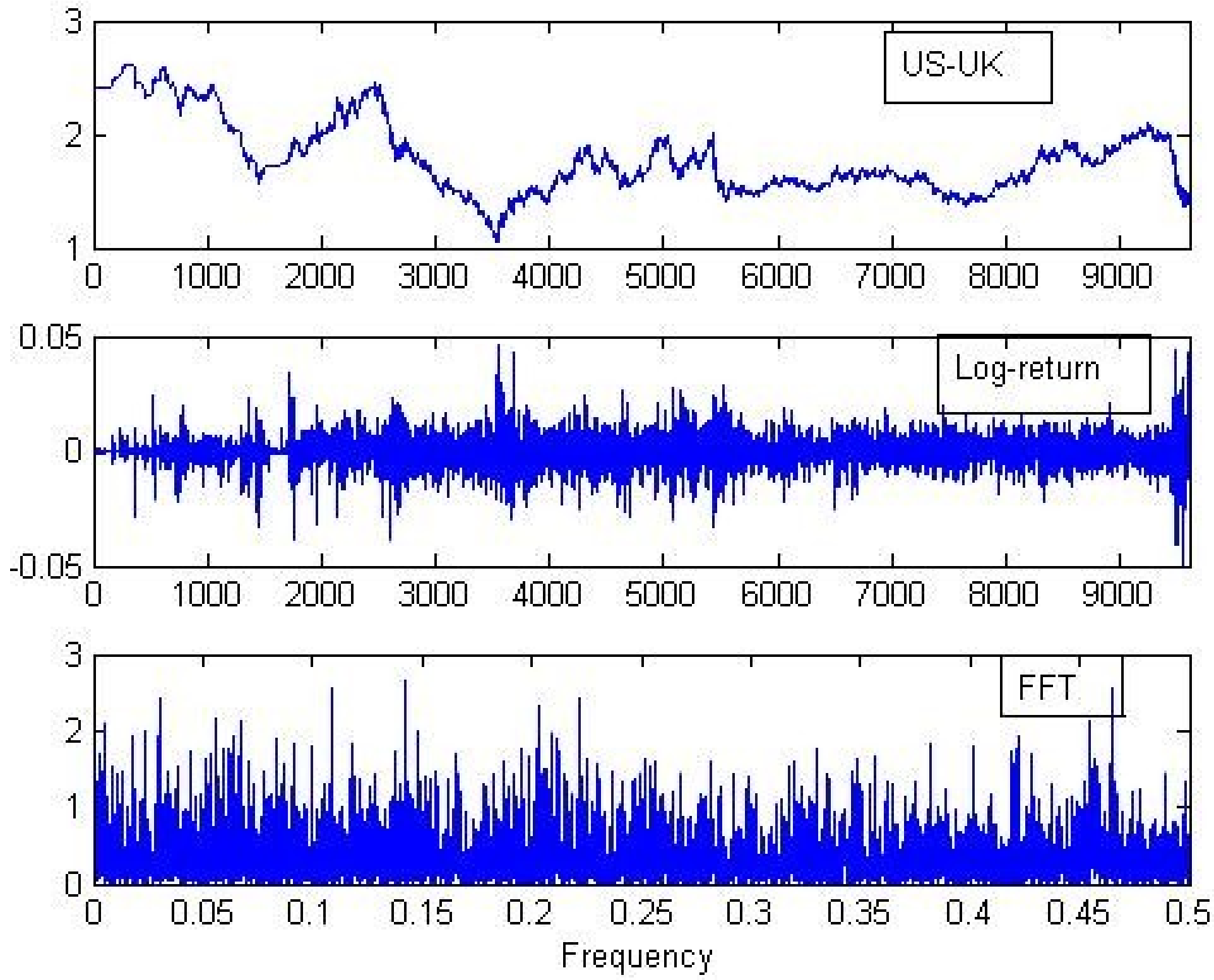

Fig2. a) Raw data of the foreign exchange rate of U.S-UK ( top most panel), log-returns of the raw data ( middle panel) and the Power Spectrum ( lowest panel) of the log-returns which is seen to be of broad band in nature. Top and middle panels ( x-axis is time and y-axis is amplitude in arbitrary units). Lowest panel ( x axis-frequency and y-axis is the power spectral amplitude).

In order to distinguish the intermittency in a time series it is usual to plot the probability distribution function of the entire time series. But with this method one cannot obtain information of the scale at which the intermittency has been observed. This disadvantage can be overcome by using the time series (coefficients) of every scale and looking at their probability distribution functions individually. We have obtained this information and fig1b ( third panel from top) for the scale of 10.

Interestingly it is seen that it is non-Gaussian almost up to scale of 80, but around scale of 20 there is a tendency to become Gaussian. But after the scale of 80 it is clearly Gaussian.

This is also verified by the fig1b ( lower most panel) where we see that the Kurtosis is almost above 3.0 up to scale of 85. Interestingly the kurtosis falls almost exponentially after scale of 40 and goes below 3.0 after 85.

We carried out a similar analysis of the US/UK cuurency and noticed similar features of chaotic dynamics in the wavelet spectrum, and also in the kurtosis as shown in figs 2a and 2b. What was interesting was the analysis of the US/INDIA foreign exchange rates figs 3a and 3b. It does not show similar features in the time-scale diagram. But at one point there was an indication of some activity, but nothing to the right or left of it. Incidentally this happens to be the year 1991, when Dr.Manmohan Singh ( the present prime minister) became the finance minister in the P.V.Narasimha Rao's Cabinet, and it was speculated that India would also go global. Not much activity is observed like the other two because Indian foreign exchange market is probably controlled by the Government to a large extent. Incidentally a similar analysis of the US/CHINA currency showed no activity at all, confirming that it is also not market driven, but controlled by the government.

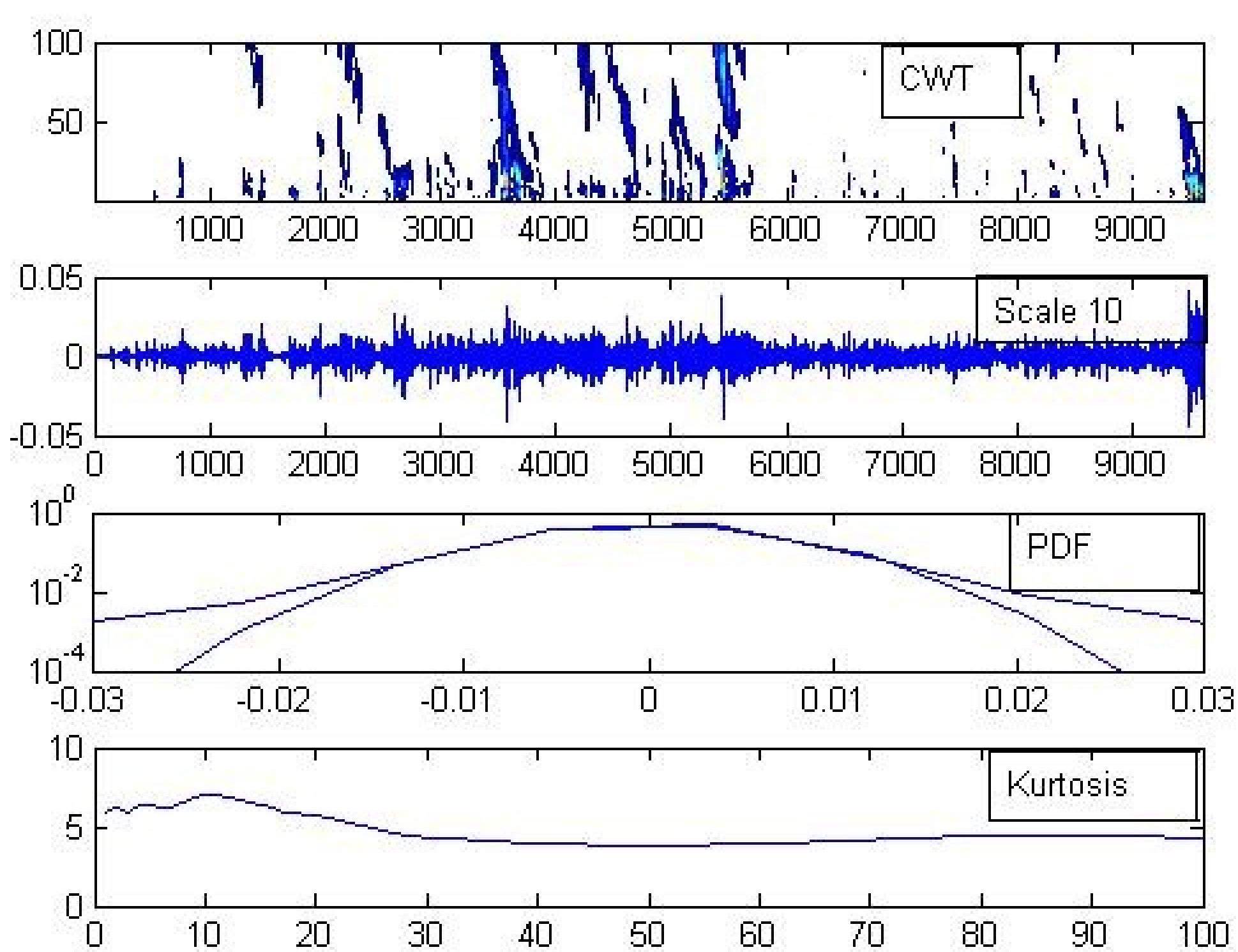

Fig2b. Time-Scale (Time along the x-axis and Scale along the y-axis) diagram obtained from the continuous wavelet transform (Top most panel), time series corresponding to scale 10 ( second from top panel), probability distribution function of the scale 10 time series data ( third panel from top) and kurtosis (lower most panel) of the scale 10 time series. Top and the second from top panels ( x-axis is time and the y-axis is the amplitude in arbitrary units). Third from top panel ( x-axis is the amplitude of the fluctuations and y-axis is the probability amplitude) and the Lowest panel ( x-axis is the scale and the y-axis is the kurtosis value)

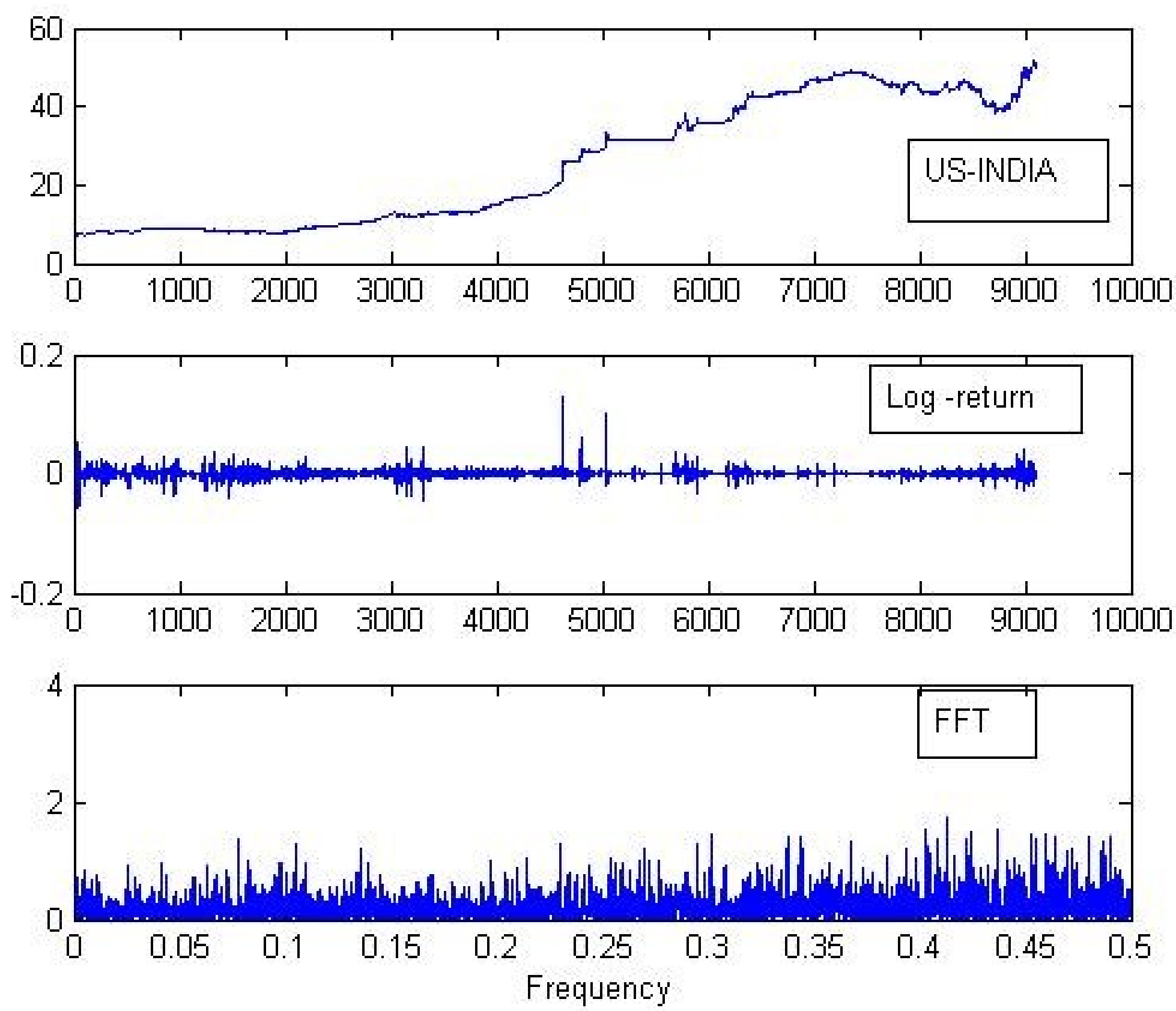

Fig3. a) Raw data of the foreign exchange rate of U.S-INDIA from 03 March 2000 to 09 March 2009, Log-returns of the time series in fig 1a ( middle panel) and the Power Spectrum of the Log-returns ( lowest panel) which is seen to be of broad band in nature. Top and middle panels ( x-axis is time and y-axis is amplitude in arbitrary units). Lowest panel ( x axis-frequency and y-axis is the power spectral amplitude).

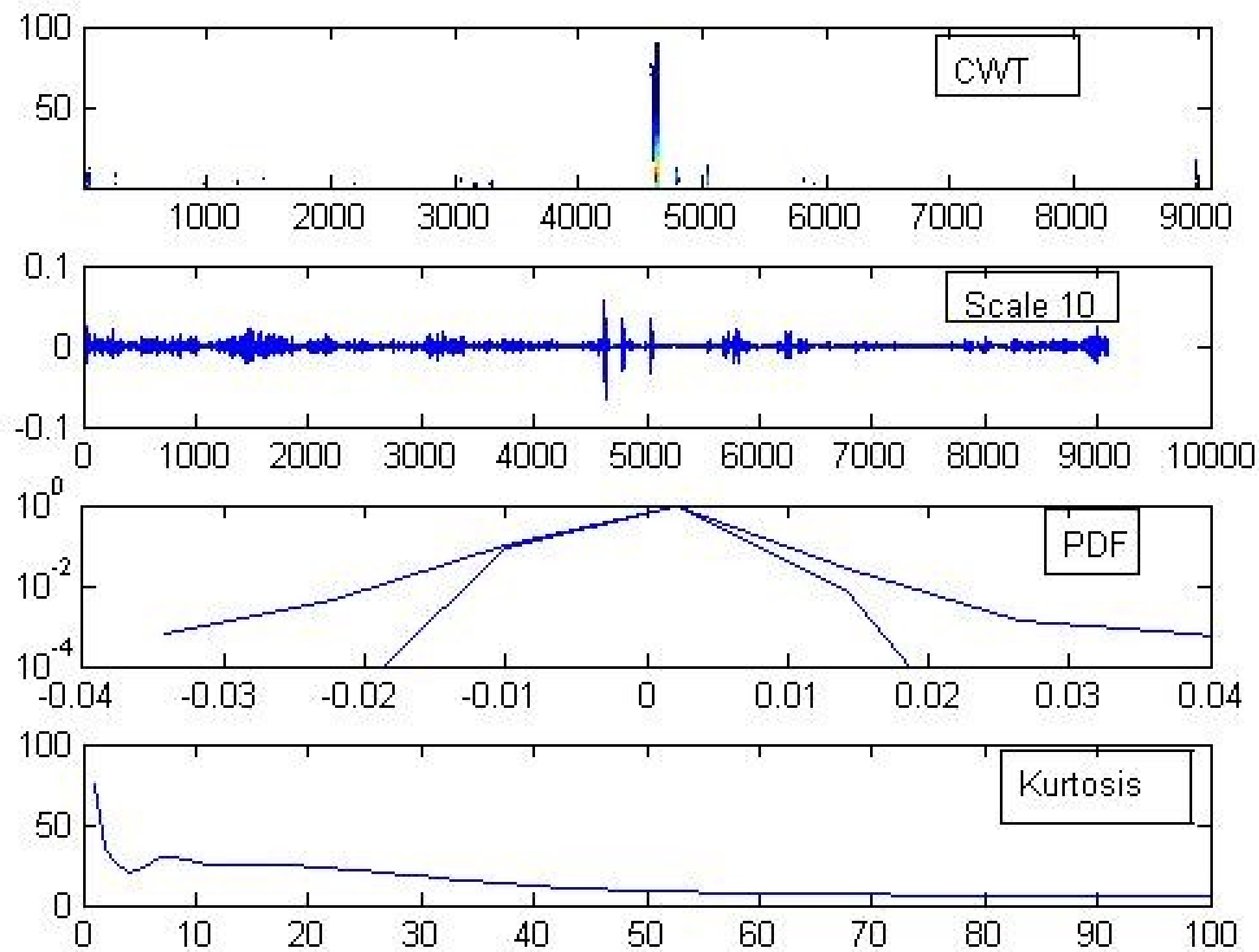

Fig3b. Time-Scale (Time along the x-axis and Scale along the y-axis) diagram obtained from the continuous wavelet transform (Top most panel), time series corresponding to scale 10 ( second from top panel), probability distribution function of the scale 10 time series data and kurtosis (lower most panel) of the scale 10 time series. Top and the second from top panels ( x-axis is time and the y-axis is the amplitude in arbitrary units). Lowest panel ( x-axis is the scale and the y-axis is the kurtosis value)

## 5. Conclusions:

We have shown using the Continuous Wavelet transforms, the presence of nonlinearity and chaos in a foreign exchange data of the U.S/EURO dollar (US/EURO) and US/UK and U/INDIA currencies. Intermittency is observed at most of the scales. Using this technique we have also shown that Indian foreign exchange rates are to a large extent controlled by the government, and not driven by the market forces almost similar to the Chinese currency which is completely controlled by the Government. We shall look into the existence or absence of similar features in other exchange rates in detail in the future.

**Acknowledgement**

We would like to thank our Director, for the support and encouragement in carrying out this work.